# Electron energy relaxation in disordered superconducting NbN films

Mariia Sidorova, Alexej Semenov, and Heinz-Wilhelm Hübers
*Institute of Optical Sensor Systems, German Aerospace Center (DLR), Rutherfordstrasse 2, 12489 Berlin, Germany*

Konstantin Ilin, and Michael Siegel,
*Institut für Mikro- und Nanoelektronische Systeme (IMS), Karlsruher Institut für Technologie (KIT), Hertzstr. 16, 76187 Karlsruhe, Germany*

Ilya Charaev
*Department of Electrical Engineering and Computer Science, Massachusetts Institute of Technology, Cambridge, Massachusetts 02139, USA*

Maria Moshkova
*National Research University Higher School of Economics, Moscow, Russia, Department of Physics, Moscow Pedagogical State University, 29 Malaya Pirogovskaya St, Moscow, 119435,Russia, and 2 LLC*

Natalia Kaurova
*Department of Physics, Moscow Pedagogical State University, 29 Malaya Pirogovskaya St, Moscow, 119435, Russia*

Gregory N. Goltsman
*Department of Physics, Moscow Pedagogical State University, 29 Malaya Pirogovskaya St, Moscow, 119435,Russia and National Research University Higher School of Economics, Moscow, Russia*

Xiaofu Zhang and Andreas Schilling
*Physics Institute, University of Zürich, Winterthurerstrasse 190, 8057 Zürich, Switzerland*

## ABSTRACT

We report on the inelastic scattering rate of electrons on phonons and relaxation of electron energy studied by means of magnetoconductance, and photoresponse, respectively, in a series of strongly disordered superconducting NbN films. The studied films with thicknesses in the range from 3 to 33 nm are characterized by different Ioffe-Regel parameters but an almost constant product $q_T l$ ($q_T$ is the wave vector of thermal phonons and $l$ is the electron mean free path). In the temperature range 14 - 30 K, the electron-phonon scattering rates obey temperature dependences close to the power law $1/\tau_{\text{e-ph}} \sim T^n$ with the exponents $n \approx 3.2 \div 3.8$. We found that in this temperature range $\tau_{\text{e-ph}}$ and $n$ of studied films vary weakly with the thickness and square resistance. At 10 K electron-phonon scattering times are in the range 11.9 - 17.5 ps. The data extracted from magnetoconductance measurements were used to describe the experimental photoresponse with the two-temperature model. For thick films, the photoresponse is reasonably well described without fitting parameters, however, for thinner films, the fit requires a smaller heat capacity of phonons. We attribute this finding to the reduced density of phonon states in thin films at low temperatures. We also show that the estimated Debye temperature in the studied NbN films is noticeably smaller than in bulk material.

## I. INTRODUCTION

Energy relaxation of nonequilibrium electrons plays an essential role in the physics of superconducting detectors. The most important relaxation processes are inelastic electron-phonon scattering and phonon escaping since they determine directly timing metrics in the performance of practical detectors. Advanced theoretical models, e.g. those of superconducting nanowire single-photon detector (SNSPD) [1] or hot-electron bolometers (HEB) [2, 3], involve not only electron-phonon scattering time, $\tau_{\text{e-ph}}$, and phonon escape time, $\tau_{\text{esc}}$, but also a few other key parameters: the ratio between electron and phonon heat capacities, $c_e/c_{\text{ph}}$, the density of electronic states at the Fermi level, and the diffusivity of electrons. Together with scattering times, they affect the energy transfer from electromagnetic radiation to electrons and further to the surrounding.

Electron-phonon scattering in bulk and clean metals is thoroughly described theoretically [4] while the acoustic mismatch model describes the escape of isotropic, three dimensional (3-d) Debye phonons from metal to dielectric through a plane boundary [5]. However, practical devices usually exploit thin and disordered superconducting films. For instance, the SNSPD with a record detection efficiency of 94% utilizes NbN film with a normal-state resistivity $5.7 \times 10^3$ Ω nm [6] that is much larger than the resistivity ≈ 550 Ω nm of crystalline stoichiometric NbN films [7]. In clean metals at low temperatures, the rate of electron phonon-scattering obeys power-law temperature dependence $1/\tau_{e-ph} \propto T^n$ with the exponent $n = 3$. The most advanced theory of electron-phonon scattering in disordered metals was developed by Sergeev and Mitin (SM) [8]. It predicts modification of the electron-phonon scattering by disorder and impurities that generally makes $n$ temperature dependent. Around a fixed temperature, $n$ depends on the degree of disorder and the kind of impurities and may have any value between 2 and 4. Furthermore, the phonon spectrum in an ultrathin film necessarily deviates from the Debye spectrum, which is commonly assumed in theories but is inherent only in bulk crystalline solids. In thin films at low temperatures, the mean free path and the wavelength of phonons become comparable or even larger than the film thickness that destroys isotropy of the phonon spectrum and reduces phonon density of states. The reduced density of states affects the strength of electron-phonon scattering and modifies its temperature dependence [9, 10] while reduced isotropy obstructs phonon escaping. The degree of phonon anisotropy depends not only on the phonon wavelength and phonon mean free path but also on the acoustic mismatch between the film and the substrate [5, 9, 11-13] via the angle of total internal reflection. Attempts to account explicitly for phonon anisotropy and reduced density of states were made phenomenologically in Refs. 5, 14 and 13 and microscopically in Ref. 12. The author of Ref. 5 introduced an effective transmission coefficient for phonons, which is an angle average of the angle-dependent transmission coefficient. The authors of Ref. 13 divided phonons into two groups and assigned them different but constant heat capacities and abilities to leave the film. The approach is referred to as the three-temperature model. Its results agree with the prediction of the microscopic model [12] where the distribution functions of electrons and phonons were computed. Another approach is called ray tracing [14]. The authors of Ref. 14 took into account breaking of Cooper pairs by phonons along with phonon scattering at non-paired electrons and traced phonons over several scattering events and reflections from the film surfaces. They showed that phonon trapping slows down the energy transfer from electrons to the substrate and that for sufficiently thin films the rate of the energy transfer does not decrease anymore with the further decrease in the film thickness.

It is important to note that the photoresponse of a detector is rather described by the relaxation time of electron energy via electron-phonon interaction, $\tau_{EP}$, which is proportional but not equal to the inelastic (single-particle) electron-phonon scattering time, $\tau_{e-ph}$, used in the theory [8]. The proportionality coefficient between these times depends on the exponent $n$ at a given temperature.

Niobium nitride, a conventional detector material, has been widely studied by means of various experimental techniques. However, significant discrepancies are present between data that have appeared in the literature over the past three decades. It is not entirely surprising. Over the years, the deposition regime of NbN films has been optimized that definitely resulted in variations of material parameters [15]. Moreover, material parameters of films, which are used for practical devices, can hardly be predicted theoretically. Knowledge of these parameters has to be acquired experimentally.

Undoubtedly, it is necessary to revise parameters of energy relaxation in modern thin superconducting NbN film. To achieve this goal, we first analyzed energy relaxation rates in NbN films reported in the literature (subsection A below) as well as the experimental techniques and models which the authors used to describe their data (subsection B). Second, employing various experimental techniques, we measured the scattering and relaxation rates and relevant parameters in two series of NbN films, which were deposited with different regimes on different substrates. In Section II, we describe these regimes and properties of specimens used for our study. Sec. III A-B describes results of transport and magnetoconductance measurements from which we extract values of electron-phonon scattering time and its temperature dependences. Sec. III C contains data on energy relaxation rates obtained with the time-domain photoresponse technique. Energy relaxation rates obtained from the photoresponse in the frequency domain are described in Section III D. We analyze our data in Section IV. Section IV A contains a comparison between our experimental data and predictions of the SM theory. Fitting experimental data with the SM theory allows us to extract the acoustic parameters of our films. With these parameters we evaluate (Sec. IV B) transmission coefficients for phonons at the film interfaces and estimate escape times for phonons in the framework of the acoustic mismatch model. We further use these times in Sec. IV C to describe photoresponse data and estimate electron and phonon heat capacities. We summarize our results in Section V.

### A. Reported electron-phonon energy relaxation rates in NbN films

In Table I we present published electron-phonon relaxation times, phonon escape times, and heat capacity ratios for NbN films with various thicknesses on different dielectric substrates. The data have been obtained by means of different experimental techniques. We also include publications reporting exclusively on transport parameters that were obtained via Hall and magnetoconductance measurements.

Table I. Reported characteristics of NbN films: $T_C$ transition temperature, $d$ film thickness, $R_{SN}$ resistance of the film square at $T > T_C$. Transport parameters of electrons in the normal state are diffusivity, $D$, elastic mean free path, $l$, and elastic scattering time, $\tau$. AMAR – Absorption of Modulated (Amplitude) sub-THz Radiation [16], 2-T – two-temperature model for electrons and phonons, MC – magnetoconductance.

| $d$, nm | $T_C$, K | $R_{SN}$, $\Omega$ | $\tau_{e-ph}(T)$, ps | $n$ | $c_e/c_{ph}(T)$ | $\tau_{esc}$, ps | Substrate | $D$, cm$^2$/s | $l$, nm | $\tau$, fs | Experimental technique & analysis | Ref. |
|---|---|---|---|---|---|---|---|---|---|---|---|---|
| 15 – 30 | 11.0 – 12.0 | 200 – 60 | 20(10)[a] | 1[a] | | | Al$_2$O$_3$ | | | | AMAR & 2-T | [17] |
| 5 | 8.5 | 450 | | | | 115 | Si/Si$_3$N$_4$ | | | | Time domain & 2-T | [18] |
| 200 - 300 | 10.3 | 34 – 20 | 7.2(10)[b] | 1.64 | | | Si/SiO$_2$ | 0.2 | | | MC | [19] |
| 2.5 -10 | | 1000 – 70 | | | | 32.5 - 130 | Al$_2$O$_3$ | | | | AMAR & 2-T | [20] |
| 20 | 8.2 | 360 | 17(8)[c] | | 0.3(8) | 160 | MgO | | | | Time-domain & 2-T | [21] |
| 7 | 11.0 | 500 | 12(7)[a] | 1.6[a] | | | Al$_2$O$_3$ | 0.4 | 0.1 | | AMAR & 2-T | [22] |
| 3.5 | 10.6 | 400-500 | 10(10)[c] | | | 38 | Al$_2$O$_3$ | | | | Time-domain & 2-T | [23] |
| 3.2 - 14.4 | 9.9 – 15.3 | 831 – 81 | | | | | Al$_2$O$_3$ | 0.51 – 0.66 | 0.58 – 0.83 | 2.16 – 3.86 | Ellipsometry | [24] |
| 12 | 14.96 | 85 | | | | | Al$_2$O$_3$ | 0.83 | 0.2 | | Hall measurements | [25] |
| 2.16 – 15 | 6.7 – 15.0 | 2377 – 107 | | | | | Al$_2$O$_3$ | | | | Tunneling spectroscopy | [26] |
| >50 | 9.99 – 16.11 | 189.2 -76.6 | | | | | MgO | | 0.207 – 0.396 | | Hall & transport measurements | [27] |
| 6 | 12.63 | 431 | | | | | | 0.544 | | | | [28] |
| 2 -20.5 | 2.6 – 15.0 | 1200 – 40 | | | | | MgO | 1.04 – 0.76 | 0.13 – 0.27 | 0.1 | MC | [29] |
| 5.2 | 11.15 | 257.7 | | 3 | | | MgO | 0.9 | | | MC | [30], [31] |

| | | | | | | | | | | | |
|---|---|---|---|---|---|---|---|---|---|---|---|
| 5.5 7 | 13.51 7.71 | 280 803 | | | | | MgO Si/SiO$_2$ | 0.92 0.47 | | | Hall & transport measurements | [32] |
| 5.5 | 7.84 | 800 | | | 0.7(8) | | Si/SiO$_2$ | 0.35 | | | Resistive thermometry & 2-T | [33] |

[a] The authors identified measured decay times of the photoresponse, $\tau_\varepsilon$, with the electron-phonon energy relaxation time, $\tau_{EP}$. The exponent $n$ relates to the temperature dependence of the photoresponse time.

[b] The inelastic electron-phonon scattering time, $\tau_{e-ph}$.

[c] The electron-phonon energy relaxation time, $\tau_{EP}$.

For all films in Table I, the resistivity, $R_{SN} d$ ($R_{SN}$ is the resistance of the film square, $d$ is the film thickness), is much larger than the resistivity ≈ 550 Ω nm of crystalline stoichiometric NbN films [7]. Although magnitudes of $\tau_{e-ph}$ are close for different films, the exponent $n$ in the temperature dependence of $\tau_{e-ph}^{-1}$ varies from 1 to 3. Since for reported films the magnitudes of inelastic (electron-phonon) scattering time, $\tau_{e-ph} \gg \tau$, where $\tau$ is the elastic scattering time of electrons, they are supposed to exhibit the phenomenon of weak electron localization [34]. The elastic mean free path, $l$, is of the order of the interatomic distance in stoichiometric NbN (0.44 nm) and the electron diffusivity in the normal state stays in the range $0.2 \leq D \leq 1$ cm$^2$/s.

Quantities related to the metal-insulator transition (Ioffe-Regel parameter $k_F l$) and the impact of disorder on the electron-phonon coupling fall in the intervals $1.5 \leq k_F l \leq 7.1$ and $0.015 \leq q_T l \leq 0.54$, respectively, where $k_F$ is the wave vector of electrons at the Fermi energy and $q_T$ is the wave vector of the thermal phonon. This classifies the films from Table I as disordered films ($q_T l \ll 1$) close to the Anderson localization limit $k_F l = 1$.

Numerical simulations have shown that the condition $k_F l \approx 1$ converts superconductor into a granular system where superconducting grains (islands) are immersed in an insulating sea and interconnected by Josephson junctions [35 - 37]. The granular morphology of thick NbN film was observed in Ref. 38. An unusually small value of the electron diffusion coefficient in thick films reported in Ref. 19 indicates either a presence of defects (vacancies or impurities) or pronounced granularity [25] that should definitely affect the inelastic scattering rate of electrons [39].

From the values of the phonon escape time shown in Table I, we estimated the product $\bar{\eta}\bar{u} = 4d/\tau_{esc}$ in the framework of the classical isotropic acoustic mismatch model [5]. Here $\bar{\eta}$ is the mean transmission of the film/substrate interface for phonons, and $\bar{u}$ is the mean sound velocity in the film (Section IV B). For the NbN/MgO interface we obtained at $\bar{\eta}\bar{u} \approx 0.5$ nm/ps, for NbN/Si$_3$N$_4$ $\bar{\eta}\bar{u} \approx 0.17$ nm/ps, and for NbN/Al$_2$O$_3$ $\bar{\eta}\bar{u} \approx 0.3 \div 0.37$ nm/ps.

### B. Electron energy relaxation: Measuring techniques and models

There are several experimental methods that allow finding the magnitude and the temperature dependence of the relaxation rate of the electron energy. We divide all of them into two distinct groups, magnetoconductance (MC) [34] and photoresponse methods, according to whether the method does not imply or does imply electron heating. Depending on whether the intensity of radiation is modulated periodically or by forming short pulses, the photoresponse is measured in the time domain or the frequency domain, respectively. Corresponding experimental techniques are usually referred to as photoresponse either in the frequency domain to amplitude-modulated radiation (FDAM) in the spectral range from sub-THz (AMAR) [16] to optics [40] or in the time domain to pulses (TDP) of radiation.

It is worth mentioning here assumptions that FDAM and TDP techniques imply. The measurements rely on a radiation-induced change in the resistance, which deals with either the concentration of free vortices (occurring in the BKT theory) or the size of a normal domain along the current path, but the response is described in terms of quasiparticles and Cooper pairs (the theory of non-equilibrium superconductivity). The FDAM and TDP techniques are applied under similar operating conditions, i.e. the current-carrying microbridge is kept at the superconducting transition and is illuminated by electromagnetic radiation with varying intensity. The intensity of incident electromagnetic radiation is modulated either periodically (FDAM) or by forming pulses (TDP). The measured quantity is the voltage drop over the current path in the microbridge that changes when the resistance of the microbridge changes. The change in resistance is caused by variation in the absorbed power of electromagnetic radiation. Absorbed energy is transferred to electrons and increases their temperature, but it doesn't change the resistance. The resistance is determined by the density of free vortices or by the size of the normal domain. It is assumed that the vortex density or the size of the domain instantly follows the electron temperature, which in its turn is controlled by the rate of absorption of electromagnetic energy and the rate of relaxation of electron energy. Crucial to these techniques is small absorbed energy that ensures the linearity of the photoresponse and exponential relaxation of the electron energy.

The MC technique allows finding the phase-breaking rate of the electron wave function for conductors in the quantum diffusive regime when electrons undergo multiple elastic (phase preserving) scattering events before the coherence (phase) of the wave function is randomized due to any inelastic (phase breaking) scattering event. In this regime, conducting electrons experience quantum interference leading to an enhanced probability for backscattering (returning to the initial position after several elastic scattering events). This quantum phenomenon is called weak localization (WL) and results in a negative correction to the normal Drude conductivity. The magnitude of this correction increases when the temperature decreases. In the presence of magnetic field, wavefunctions (corresponding to clockwise and counterclockwise trajectories along the same loop) acquire different phase shifts and interfere at the initial point destructively. Hence, magnetic field destroys the enhanced backscattering. Since the maximum length of trajectories contributing to WL is limited to the inelastic scattering length, this length can be evaluated by measuring the field that suppresses WL. Corresponding phase-breaking rate is a sum of rates of all inelastic scattering processes. Electron-phonon scattering dominates in phase-breaking at temperatures well above $T_C$. In the vicinity of $T_C$, phase-breaking and the correction to the conductivity are additionally affected by superconducting fluctuations (see details in Sec. III B).

With respect to *e-ph* scattering, the phase-breaking rate is identical to the inelastic single-particle *e-ph* scattering rate at the Fermi energy [41], $\tau_{e-ph}^{-1}$, which is considered in the SM theory. Because the photoresponse technique implies electron heating, the *e-ph* energy relaxation rate, $\tau_{EP}^{-1}$, extracted by means of this technique differs from the single-particle scattering rate. The energy relaxation rate is just an average of the single-particle scattering rate over the range of electron states $\sim k_B T$. The relationship between these rates was obtained in [42] as follows

$$\tau_{EP}^{-1} = \frac{3(n+2)\Gamma(n+2)\varsigma(n+2)}{2\pi^2(2-2^{1-n})\Gamma(n)\varsigma(n)} \tau_{e-ph}^{-1}, \qquad (1)$$

where $\Gamma(n)$ is the gamma function, $n$ is the exponent in the temperature dependence of the scattering rate $\tau_{e-ph}^{-1} \propto T^n$, and $\varsigma(n)$ is the Riemann zeta function.

Data analysis in AMAR, FDAM, and TDP methods is based on the two-temperature (2-T) model [43], which is an extension of the Rothwarf-Taylor model [44] for temperatures close to $T_C$. The 2-T model implies that electrons and phonons are instantly in the internal equilibrium and are described by their equilibrium distribution functions with two different effective temperatures, which are slightly larger than the ambient temperature. The evolution of the effective temperatures caused by external excitation is described by a system of two coupled time-dependent equations. It is assumed that the rate of the decay of the excess phonon energy is a sum of rates $\tau_{esc}^{-1}$ and $\tau_{PE}^{-1}$ associated with escaping of phonons from the film into the substrate and with phonon-electron scattering, respectively. The 2-T model accounts for phonon trapping, i.e. the angle of total internal reflection of phonons at the film/substrate interface, $\theta_{max}$, by assigning to all phonons the same escape rate $\tau_{esc}^{-1}$. This mean escape rate is less than the escape rate for phonons hitting the interface at angles $\theta < \theta_{max}$. The principle of detailed balance [44] requires that in equilibrium the energy flow from electrons to phonons equals the backward flow. This equality relates the heat capacity ratio to the ratio of energy relaxation times for electron and phonons as $c_e/c_{ph} = \tau_{EP}/\tau_{PE}$ [43].

## II. SPECIMENS AND PARAMETERS

We studied thin NbN films with different thicknesses and different degrees of disorder. The specimens are listed in Table II. Films of the M-series (2559, A853, A854, and A855) were magnetron-sputtered onto silicon substrates on top of a thermally prepared layer of silicon oxide with a thickness of 250 nm. Magnetoconductance measurements (Sec. III B) were carried out with non-structured approximately squared 1x1 cm$^2$ NbN films. TDP measurements (Sec. III C) were carried out with the same films, which were shaped in the form of microbridges. The lengths of microbridges varied from 3.6 to 7 μm, and the widths from 0.615 to 0.69 μm. The sizes were chosen in order to match the normal square resistance of each microbridge to the electrical impedance of the readout circuit ($Z_0$ = 50 Ω). FDAM measurements (Sec. III D) were carried out with NbN microbridges of K-series on sapphire substrates (K-1 – K-9). They had thicknesses in the range from 3.2 to 33.2 nm. Films of K-series were also magnetron-sputtered. The sputtering regime was optimized for the largest $T_C$. The fabrication process of these K-films is described in detail in Refs. 40 and 15. Measurements of the density of electronic states, transition temperature, and diffusivity are described below in Section III A.

Table II. Parameters of studied NbN films. $R^{300K}/R^{20K}$ is the ratio of the resistances at 300 and 20 K, $N(0)$ is the total density of states for electrons at the Fermi energy.

| Sample | $d$ (nm) | $T_C$ (K) | $R_{SN}$ (Ω/sq) | $D$ ($10^{-4}$ m$^2$/s) | $N(0)$ (eV$^{-1}$ m$^{-3}$) | $R^{300K}/R^{20K}$ |
|---|---|---|---|---|---|---|
| M-2259 | 5.0 | 10.74 | 529.5 | 0.474 | 4.98×10$^{28}$ | 0.793 |
| M-A853 | 6.4 | 8.35 | 954.0 | 0.339 | 3.02×10$^{28}$ | 0.709 |
| M-A854 | 7.5 | 10.84 | 387.9 | 0.453 | 4.74×10$^{28}$ | 0.809 |
| M-A855 | 9.5 | 10.94 | 330.6 | 0.418 | 4.75×10$^{28}$ | 0.788 |
| K-1 | 3.2 | 12.70 | | | | 0.83 |
| K-2 | 4.2 | 12.90 | 450 | 0.53 | 6.5×10$^{28}$ | 0.90 |
| K-3 | 5.8 | 14.60 | | | | |
| K-4 | 7.5 | 14.80 | | | | 1.000 |
| K-5 | 8.6 | 15.35 | | | | |
| K-6 | 9.9 | 10.80 | 90 | | | 1.025 |
| K-7 | 14.9 | 16.00 | | | | |
| K-8 | 21.6 | 16.35 | | | | 1.023 |
| K-9 | 33.2 | 16.35 | | | | |

As seen from Table II, the films of similar thicknesses M-2259, M-A853, and K-2 are characterized by different degrees of disorder in terms of the Ioffe-Regel criterion [45]. For the film K-2, $k_Fl = 3Dm_e/\hbar \approx 1.37$, while for films of the M-series $k_Fl$ varies from 0.88 to 1.22 (values of the electron mean free path are evaluated in Sec. IV A). Furthermore, the diffusion coefficient, the transition temperature, and the total density of electronic states, $N(0)$, of the film K2 are larger while the square resistance, $R_{SN}$, is lower than these parameters of the films from the M-series. The numbers indicate [46, Chapter 3 in Ref. 15] that the composition NbN$_x$ of the film K-2 is characterized by $x \approx 1.04$ and a higher content of niobium than the composition of films of the M-series with x ≈ 1.18. Stoichiometric composition corresponds to $x = 1$. Films M-2259, -A854, -A855 have close values of the electron diffusion coefficient, electron density of states, and transition temperature. These parameters are noticeably smaller for the film M-A853, while its square resistance and resistivity ($R_{SN} d$) are much larger as compared to others. Correspondingly, among films of the M-series, the film M-A853 has the largest degree of disorder, $k_Fl$=0.88. It is close to the superconductor-insulator transition [35] and may additionally have an enhanced degree of granularity (see Sec. IV A). We have to note here that the parameters of the films of the K-series are close to those reported for similar films in Ref. 24.

## III. EXPERIMENT AND RESULTS

### A. DC transport and superconducting properties

Transport measurements were carried out by the standard four-probe technique in a Physical Property Measurement System (PPMS) manufactured by Quantum Design. Applied bias currents were less than 100 μA. The square resistance, $R_S$, was measured with the van der Pauw method that eliminates the effect of the planar geometry for 2-d specimens. In Fig. 1 we show $R_S(T)$ dependences for four NbN films of M-series with different thicknesses. As seen in the inset, for each film $R_S$ increases with the decrease in the temperature from 300 K down to approximately 20 K that is most likely due to Anderson localization. At lower temperatures, the $R_S(T)$ dependences flatten, the square resistance of each film reaches a plateau and then goes down to zero value within a finite transition region caused by superconducting fluctuations.

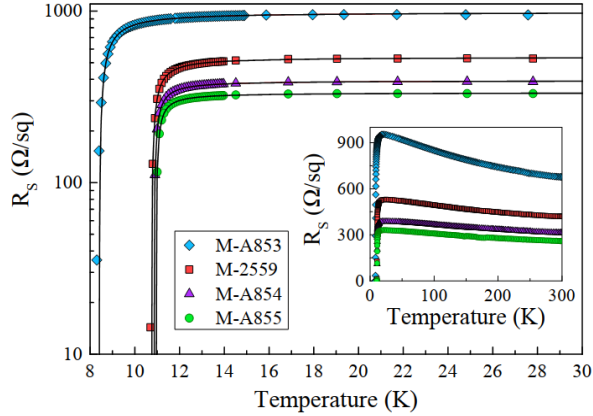

Fig.1. Temperature variation of the square resistance for four exemplary films with different thicknesses around their superconducting transitions. Solid lines represent the best fits obtained with Eq. (2) in the vicinity of $T_C$ and extrapolated to 30 K. The inset shows resistances in the broader temperature range up to 300 K.

We fit our experimental $R_S(T)$ data with the theory of fluctuation conductivity of Aslamazov and Larkin (AL) [47] and Maki and Thompson (MT) [48, 49]. For two dimensional films, the theory predicts

$$R_S(T) = \frac{R_{SN}}{1 + R_{SN} Y \frac{1}{16} \frac{e^2}{\hbar} \left(\frac{T_C}{T - T_C}\right)}, \qquad (2)$$

where $\hbar$ is the reduced Planck constant, $e$ is the elementary charge, $T_C$ is the BCS mean-field transition temperature, and $R_{SN}$ is the normal-state square resistance at a temperature right above the superconducting transition.

We used $T_C$, $R_{SN}$, and $Y$ as fitting parameters to fit experimental data in a very narrow temperature range for which the inequality $\ln(T/T_C) \ll 1$ holds. Best fit values of $T_C$ and $R_{SN}$ are listed in Table II. Normal state resistances extracted from the fits are slightly larger than the measured at 20 K. The fitting parameter $Y$ varies between 1.9 – 2.6 for all films. The presence of two types of excitation: topological (magnetic vortices) and electronic (quasiparticles) complicates the definition of the superconducting transition temperature in two dimensional (2-d) films. It turns out that highly resistive 2-d superconducting films exhibit two transition temperatures. One of them, $T_{BKT}$ (Berezinskii-Kosterlitz-Thouless), controls unbinding of vortex-antivortex pairs that provides emergence and an exponential rise of the resistance with increasing density of free vortices. The other controls the energy gap. It is known as the mean-field transition temperature and doesn't cause the emergence of the resistance. In Ref. 28 it was reported that for NbN film with $R_{SN} = 431$ Ω/sq these temperatures are related as $T_{BKT} = 0.85 T_C$. Anyway, right above the superconducting transition, our experimental $R_S(T)$ dependences are well described by AL and MT fluctuations.

Applying external magnetic field perpendicularly to the film surface, we measured $R_S(T)$ dependences for a set of magnetic fields. The preset field was taken as the second critical field at the temperature which corresponds to the midpoint of the transition, i.e. $R_S = R_{SN}/2$. This procedure gives the second critical magnetic field, $B_{C2}$, as a function of temperature for the temperature range below $T_C$. For all our films, we found almost linear behavior of $B_{C2}$ vs $T$ in the range $T_C/2 < T < T_C$ and used the slope for computing the electron diffusion coefficient as [50]

$$D = \frac{4 k_B}{\pi e} \left(\frac{dB_{c2}}{dT}\right)^{-1}. \tag{3}$$

The values of $D$ are listed in Table II along with the total electron density of states at the Fermi energy, $N(0)$, which we computed using Einstein relation $N(0) = 1/(e^2 R_{SN} d\, D)$.

### B. Magnetoconductance

Films of M-series represent disordered 2-d systems suitable for the MC method. We use the same PPMS apparatus as for DC measurements, to acquire square resistance $R_S(B,T)$ at different fixed temperatures in the range from $T_C$ to $3T_C$ by varying magnetic field in the range from 0 to 9 T. The dimensionless change in the conductance per sample square induced by the field at the fixed temperature $T$ was determined according to

$$\delta\sigma(B,T) = \frac{2\pi^2 \hbar}{e^2}\left[\frac{1}{R_S(B,T)} - \frac{1}{R_S(0,T)}\right].$$

Experimental data are shown in Fig. 2. Since dependences $\delta\sigma(B,T)$ are monotonous and look pretty similar for all studied specimens, we plot in Fig. 2 data for only one representative film.

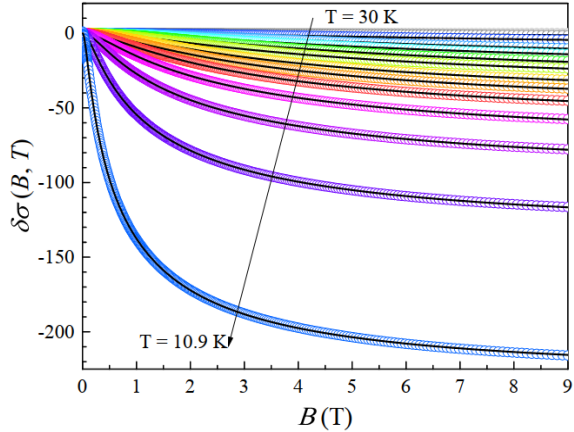

Fig.2. Field induced change in the conductance (Eq.(3)) for the film M-2259 vs magnetic field. Different colors correspond to different temperatures. Solid black curves are fits with Eqs. (4) – (7).

Contribution to the magnetoconductance $\delta\sigma(B,T)$ due to the effect of weak localization has the form [51]

$$\delta\sigma^{WL}(B,T) = \frac{3}{2}\psi\left(\frac{1}{2} + \frac{B_2}{B}\right) - \frac{1}{2}\psi\left(\frac{1}{2} + \frac{B_i}{B}\right) + \frac{3}{2}\ln\left(\frac{B}{B_2}\right) - \frac{1}{2}\ln\left(\frac{B}{B_i}\right), \tag{4}$$

where $\psi(x)$ is the di-gamma function, $B_i = \hbar/(4eD\tau_\varphi)$ is the inelastic magnetic fields, $\tau_\varphi$ is the phase-breaking time, $B_2 = B_i + \frac{4}{3}B_{s.o.}$, $B_{s.o.} = \hbar/(4eD\tau_{s.o.})$, and $\tau_{s.o.}$ is the spin-orbit scattering time. The WL correction provides a positive contribution to $\delta\sigma(B,T)$; its magnitude increases with the increase of magnetic field.

Superconducting fluctuations (stochastic formation of Cooper pair) decrease the time that electrons remain normal, i.e. decrease their mean concentration and increase effective conductance. This causes the broadening of the superconducting $R_S(T)$ transition at $T > T_C$. Since the increase in conductivity due to fluctuations is reduced by the external magnetic field, fluctuations provide a negative contribution to $\delta\sigma(B,T)$. The effect is commonly denoted as the Aslamazov-Larkin correction to magnetoconductance. In the 2-d limit and in the immediate vicinity of $T_C$, where the AL contribution dominates $\delta\sigma(B,T)$, it has the form [47, 52]

$$\delta\sigma^{AL}(B,T) = -\frac{\pi^2}{8\ln\left(\frac{T}{T_C}\right)}\left[8\left(\frac{\tilde{B}_C}{B}\right)^2\left\{\psi\left(\frac{1}{2} + \frac{\tilde{B}_C}{B}\right) - \psi\left(1 + \frac{\tilde{B}_C}{B}\right) + \frac{B}{2\tilde{B}_C}\right\} - 1\right]. \qquad (5)$$

Here $\tilde{B}_C$ is the characteristic field defined by the relation $\tilde{B}_C = C\frac{k_BT}{\pi eD}\ln\left(\frac{T}{T_C}\right)$. In different publications, the numerical factor $C$ were assigned values from 2 to 6 [53, 54-57].

Maki-Thompson correction to magnetoconductance accounts for stochastic, for a time shorter than $\tau_\varphi$, pairing of two electrons, which are about to simultaneously (coherently) scatter at the same scattering center. Paring eliminates scattering that effectively increases the electron mean free path and weakens the effect of localization. Since in the 2-d limit localization causes correction to conductance $\delta\sigma \propto -\ln(L_\varphi/l)$ ($L_\varphi$ is the phase breaking length), such events give a negative contribution to $\delta\sigma(B,T)$. The MT correction [48, 49] for the 2-d limit was elaborated by Larkin [58]. The contribution is given by

$$\delta\sigma^{MT}_{(*)}(B,T) = -\beta_L(T)\left[\psi\left(\frac{1}{2} + \frac{B_i}{B}\right) + \ln\left(\frac{B}{B_i}\right)\right], \qquad (6)$$

where $\beta_L(T) = \pi^2/[4\ln(T/T_C)]$ at $\ln(T/T_C) \ll 1$. The MT contribution was further modified by Lopes dos Santos and Abrahams (LSA) for the temperatures close to $T_C$ ($\ln(T/T_C) \ll 1$) [59] as

$$\delta\sigma^{MT}_{(mod)}(B,T) = -\beta_{LSA}(T,\delta)\left[\psi\left(\frac{1}{2} + \frac{B_i}{B}\right) - \psi\left(\frac{1}{2} + \frac{\tilde{B}_C}{B}\right) - \ln\left(\frac{\tilde{B}_C}{B_i}\right)\right], \qquad (7)$$

where $\beta_{LSA}(T,\delta) = \pi^2/\{4[\ln(T/T_C) - \delta]\}$ and $\delta = \pi eDB_i/(2k_BT)$ is the MT pair-breaking parameter [60-62].

At the first stage, we fit experimental data in Fig. 2 with a sum of WL contribution and contributions due to MT and AL fluctuations as $\sigma(B,T) = \delta\sigma^{WL}(B,T) + \delta\sigma^{AL}(B,T) + \delta\sigma^{MT}(B,T)$. Although $\beta_{LSA}$ is defined via $B_i$, we used $\beta_{LSA}$, $B_{s.o.}$, and $B_i$ as independent fitting parameters. This was done in order to account for possible contribution from DOS (density of electronic states) fluctuations [63, 64]. From the best fit values of $B_i$ we found $\tau_\varphi = \hbar/(4eDB_i)$ for a set of fixed temperatures. The electron diffusion coefficient was taken from the magneto-transport measurements. In Fig. 3a we show $1/\tau_\varphi$ as a function of temperature.

The total phase-breaking rate extracted from magnetoconductance measurements is the sum of rates affiliated with independent inelastic interactions in which electrons are involved. They are electron-electron interaction (*e-e*) [65], electron-phonon interaction (*e-ph*) and electron-fluctuation interaction (*e-fl*) [66]. Although phase breaking occurs due to inelastic scattering, the phase-breaking rate may differ from the inelastic scattering rate. The difference is most pronounced for scattering processes in which the change in the electron energy is smaller than thermal energy. The phase-breaking rate due to electron-electron scattering is dominated at low temperatures by Nyquist noise, i.e. scattering with small energy transfer. In the two dimensional case, both rates have the same temperature dependence $\propto T$ but different magnitudes [65, 67]. For our quasi two-dimensional films, the phase-breaking rate is one to two orders of magnitude larger than the electron-electron scattering rate with small energy transfer. At temperatures larger than the cross-over temperature $\hbar k_B^{-1}\tau^{-1}$ [68], where $\tau$ is the elastic scattering time, the phase-breaking rate is dominated by Landau scattering, i.e. the electron-electron scattering with large energy transfer. In this latter case the phase-breaking rate equals the electron-electron scattering rate $\tau_{e-e}^{-1} \propto T^2 \ln(T^{-1})$ [69]. For our strongly disordered NbN films $\tau < 5$ fs [24] and the crossover is expected to occur at temperatures larger than $10^3$ K. Furthermore, at the upper boundary of our temperature range the magnitudes of $\tau_\varphi^{-1}$ due to Nyquist noise and due to Landau scattering differ by two orders of magnitude. Therefore in Eq. (8) we retain only one contribution to the phase-breaking rate which is affiliated with Nyquist noise. For *e-ph* interaction, the electron-phonon scattering rate and the phase-breaking rate due to this interaction are identical [41].

Phase breaking via electron-fluctuations is associated with the loss of the electron energy and phase coherence due to recombination of electrons into superconducting pairs [66]. Hence, the total phase-breaking rate is given by $\tau_\varphi^{-1} = \tau_{(e-e)}^{-1} + \tau_{e-ph}^{-1} + \tau_{(e-fl)}^{-1}$. Brackets are used to stress the difference between electron scattering rates and respective contributions to the phase-breaking rate. The contributing rates are:

$$\begin{cases} \tau_{(e-e)}^{-1} = \frac{k_B T}{\hbar} \frac{1}{2C_1} \ln(C_1) \\ \tau_{e-ph}^{-1} = C_2 (T/T_C)^n \\ \tau_{(e-fl)}^{-1} = \frac{k_B T}{\hbar} \frac{1}{2C_1} \frac{2\ln(2)}{\ln(T/T_C) + C_3} \end{cases} \quad (8)$$

where $C_1 = \pi\hbar/R_{SN}e^2$ and $C_3 = 4\ln(2)/\left[\sqrt{\ln(C_1)^2 + 128C_1/\pi} - \ln(C_1)\right]$ [66].

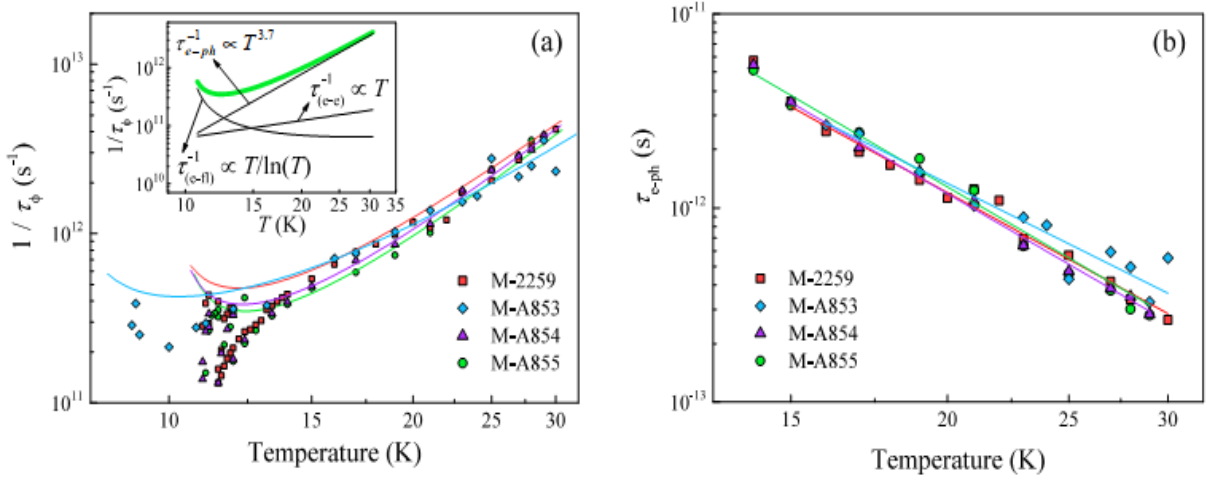

Fig.3. (Color online) (a) Phase-breaking rate vs temperature (symbols) extracted from magnetoconductance measurements in the double logarithmic scale. Solid lines are fits made with the sum of all three terms in Eq. (8). The inset shows the best fitting curve for the film M-A855 (thick green line) and separately all three terms (thin black lines). (b) *e-ph* scattering time vs temperature extracted from magnetoconductance measurements (symbols) in the double logarithmic scale. Solid lines are fits obtained with the second term in Eq. (8). Fitting parameters are listed in Table III.

At the second stage, we fit the temperature dependence of the experimental phase-breaking rate $\tau_\varphi^{-1}(T)$ with the sum of contributions (Eq. 8) of different scattering processes. We assume arbitrary but temperature-independent $n$ and use $C_2$ and $n$ as fitting parameters. As has been noted in other publications [30, 31], it is not possible to fit experimental data for $\tau_\varphi^{-1}(T)$ at temperatures close to the superconducting transition. The reason is not clear and goes beyond the scope of our study. To circumvent the problem, we included in the fitting procedure only the data obtained at temperatures above 14 K. The result is shown in Fig. 3a for four different films. The best fit values of $n$ are listed in Table III and $C_2$ are $10.8\times10^{10}$, $4.6\times10^{10}$, $8.48\times10^{10}$, and $8.02\times10^{10}$ s$^{-1}$ for M-2259, M-A853, M-A854,

and M-A855, respectively. Inset in Fig. 3a shows the total rate $1/\tau_\varphi(T)$ for the film A855 (thick green curve) and all three contributions separately (black thin curves). It is clearly seen that at $T \gg T_C$ the term $\tau_{e-ph}^{-1}$ dominates and defines both the temperature dependence and the magnitude of the total scattering rate. Contrarily, close to $T_C$ the term $\tau_{(e-fl)}^{-1}$ dominates and controls the upturn in the $\tau_\varphi^{-1}(T)$ dependence. Subtracting (e-e) and (e-fl) contributions (Eq. 8) from the experimental phase-breaking rate, we obtained electron-phonon scattering rate $\tau_{e-ph}^{-1}$ for each specimen. Fig. 3b shows corresponding values of $\tau_{e-ph}$ in the temperature range from 14 to 30 K. Solid lines represent temperature dependences predicted by the second term in Eq. (8) with the best fit values of $C_2$ and $n$ for each specimen. The values of $\tau_{e-ph}$ extrapolated to 10 K according to Eq. (8) are listed in Table III.

Table III. Parameters of NbN on Si/SiO$_2$ substrate. Heat capacity ratios refer to the transition temperatures.

| sample | $d$ (nm) | Transport measurements (Sec. III A) | | | | MC technique (Sec. III B) | | TD technique (Sec. III C) | | |
|---|---|---|---|---|---|---|---|---|---|---|
| | | $T_C$ (K) | $R_{SN}$ (Ω/sq) | $D$ ($10^{-4}$ m$^2$/s) | $N(0)$ (eV$^{-1}$ m$^{-3}$) | $\tau_{e-ph}$[a] (10K) (ps) | $n$ $\tau_{e-ph}^{-1} \sim T^n$ | $\tau_{EP}$[b] ($T_C$) (ps) | $\tau_{esc}$[c] (ps) | $c_e / c_{ph}$ ($T_C$) |
| M-2259 | 5.0 | 10.74 | 529.5 | 0.474 | 4.98×10$^{28}$ | 11.9 | 3.53 | 1.4 | 25.9 | 0.83 ± 0.18 |
| M-A853 | 6.4 | 8.35 | 954.0 | 0.339 | 3.02×10$^{28}$ | 12.4 | 3.21 | 4.2 | 39.0 | 0.25 ± 0.03 |
| M-A854 | 7.5 | 10.84 | 387.9 | 0.453 | 4.74×10$^{28}$ | 15.9 | 3.75 | 1.5 | 39.0 | 0.35 ± 0.05 |
| M-A855 | 9.5 | 10.94 | 330.6 | 0.418 | 4.75×10$^{28}$ | 17.5 | 3.77 | 1.6 | 51.3 | 0.11 ± 0.03 |

[a] The phase-breaking time due to *e-ph* scattering is identical with the single-particle *e-ph* scattering time [41].
[b] The *e-ph* energy relaxation time (Eq. (1)).
[c] The phonon escape time is derived in Sec. IV B.

Data in Table III show that the magnitude of $\tau_{e-ph}$ and the exponent $n$ in its temperature dependence extracted with the MC technique slightly vary with the film thickness and the sheet resistance.

As it was mentioned above, the *e-ph* scattering time obtained by means of the MC technique can be directly compared with the *e-ph* scattering time predicted by the SM theory. We apply this theory to independently fit experimental data shown in Fig. 3b and to extract acoustic parameters for each film of the M-series. The SM theory predicts *e-ph* scattering time $\tau_{e-ph}(u, \rho, l, N(0), T)$ as a function of temperature and four material parameters. Among them, $u$ and $\rho$ are the sound velocity and the mass density, respectively. Mathematical details are presented in Sec. IV A along with the best fit values of $u$, $\rho$, and $l$ (Table IV). The variation of the exponents $n$ obtained with the SM theory in the temperature range of Fig. 3b is less than 1 percent while the mean values coincide with the best fit values obtained with Eq. 8. We, therefore, do not explicitly show the best fit curves obtained in the framework of the SM theory.

We use acoustic parameters, along with the phonon velocities in substrates, to compute transmission coefficients and escape times for phonons at studied film-substrate interfaces (Section IV B). Thus obtained escape times are used in the next two sections as seed values for modeling the photoresponse of our films in the frameworks of the 2-T model.

### C. Photoresponse in the time-domain

We studied photoresponse of superconducting microbridges to sub-picosecond pulses with a wavelength of 800 nm at a repetition rate of 80 MHz. Microbridges were made from films of the M-series listed in Table II. They were mounted in a continuous flow cryostat with optical access through a quartz window. Microbridges were kept in the resistive state at an ambient temperature $T \geq T_C$ and biased by small dc current. The photoresponse of the bridge in the form of a voltage transient was amplified within a limited frequency band 0.1 - 5 GHz and recorded with a sampling scope. Fig. 4 shows voltage transients recorded by the oscilloscope. Transients delivered by microbridges with different thicknesses look similar. They all exhibit identical rising times. Obviously, this time is limited to the bandwidth of the readout, while the falling parts of the transients still contain valuable information. Impedance matching between the microbridge and the readout is not perfect. Mismatch causes multiple reflections (signal ringing), which are poorly seen in Fig. 4a. In Fig. 4b, we plot the transients in the logarithmic scale that emphasizes the ringing. We found the ringing period of approximately 250 ps that corresponds to the propagation time of the transient over 2.5 cm electrical path between the microbridge and the first SMA connector at the microbridge holder.

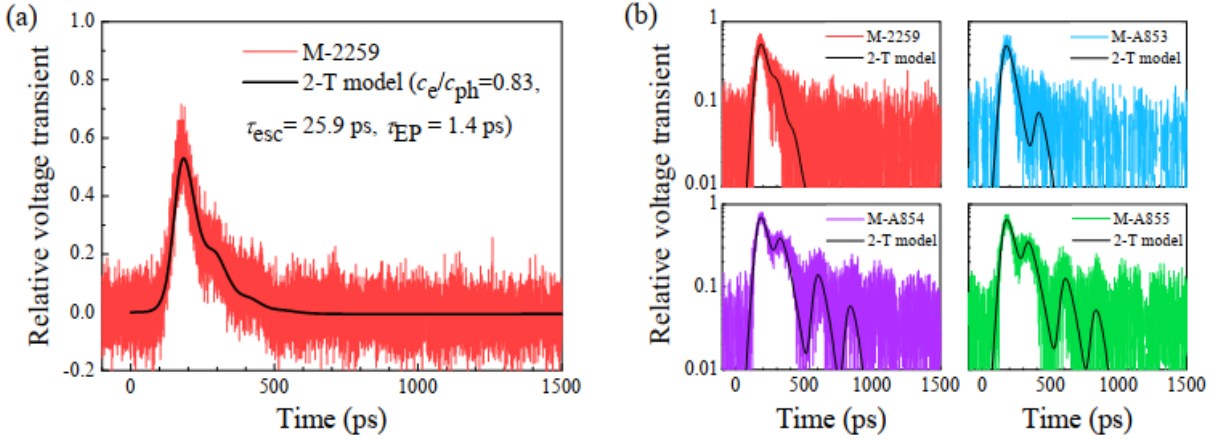

Fig. 4. (Colors online) (a) Voltage transient for the microbridge M-2259 in the linear scale. (b) Voltage transients for NbN microbridges with four different thicknesses in the semi-logarithmic scale. Black curves are best fits according to Eqs. (9-13) with parameters: for M-2259 $c_e/c_{ph}$ = 0.83 ± 0.18, $\tau_{esc}$ = 25.9 ps; for M-A853 $c_e/c_{ph}$ = 0.25 ± 0.03, $\tau_{esc}$ = 39 ps; for M-A854 $c_e/c_{ph}$ = 0.35 ± 0.05, $\tau_{esc}$ = 39 ps; and for M-A855 $c_e/c_{ph}$ = 0.11 ± 0.03, $\tau_{esc}$ = 51.3 ps, for each bridge $\tau_{EP}$ was fixed at the value obtained for original film from MC measurements and further averaged according to Eq. (1). Legends specify film from Table II.

In order to extract intrinsic relaxation times, we apply equations of the two-temperature model with pulse excitation [21]. This approach is commonly used to describe the non-equilibrium state created by an optical pulse in a resistive or superconducting film [23]. With dimensionless units for time and energy, equations of the 2-T model look as follows:

$$\begin{cases} \frac{dT_e(\xi)}{d\xi} = -\Gamma_1(T_e(\xi) - T_{ph}(\xi)) + \frac{\tau_0}{d\,c_e}P_{RF}(\xi) + \frac{\tau_0}{c_e}P_{DC} \\ \frac{dT_{ph}(\xi)}{d\xi} = \Gamma_2(T_e(\xi) - T_{ph}(\xi)) - \Gamma_3(T_{ph}(\xi) - T_0) \end{cases}, \quad (9)$$

where $T_e$ and $T_{ph}$ - are temperatures of the electron and phonon subsystems, $T_0$ – is the bath temperature, $\Gamma_1 = \frac{\tau_0}{\tau_{EP}}$, $\Gamma_2 = \Gamma_1 \frac{c_e}{c_{ph}}$, $\Gamma_3 = \frac{\tau_0}{\tau_{esc}}$, $\xi = t/\tau_0$ is the dimensionless time $P_{RF}(t) = m^3\xi^2 e^{-m\xi}E_0/\tau_0$ - analytical expression describing instantaneous power of the excitation pulse. For $m$ = 3.4, $\tau_0$ represents the full width at half maximum, and $E_0$ is the total pulse energy absorbed by the unit area of the film. $P_{DC}$ is the Joule power dissipated in the unit volume of the film. The magnitude of $P_{DC}$ was extremely small in our measurements, and therefore we neglected it. Solving Eqs. (9), we obtain time-dependent $T_e(\xi)$ and $T_{ph}(\xi)$ in the form.

$$\frac{T_e(\xi)-T_0}{T_0} = A_1 \frac{(\chi_1+\Gamma_2+\Gamma_3)}{\Gamma_2}e^{\chi_1\xi} + A_2 \frac{(\chi_2+\Gamma_2+\Gamma_3)}{\Gamma_2}e^{\chi_2\xi} + Q_1(\xi)e^{-m\xi},$$

$$\frac{T_{\text{ph}}(\xi)-T_0}{T_0} = A_1 e^{\chi_1 \xi} + A_2 e^{\chi_2 \xi} + Q_2(\xi) e^{-m\xi} \tag{10}$$

with parameters given by

$$\chi_{1,2} = -\frac{1}{2}\left(\sum_i^3 \Gamma_i \mp \sqrt{(\sum_i^3 \Gamma_i)^2 - 4\Gamma_1 \Gamma_3}\right),$$

$$A_{1,2} = \pm \frac{\Gamma_2 E_0 m^3}{d\, c_e T_0} \frac{1}{(\chi_1-\chi_2)(m+\chi_{1,2})^3},$$

$$Q_2 = \frac{\Gamma_2 E_0 m^3}{d\, c_e T_0}(a\,\xi^2 + b\,\xi + c), \tag{11}$$

$$Q_1 = \frac{E_0 m^3}{d\, c_e T_0}\left((\Gamma_2 + \Gamma_3 - m)(a\,\xi^2 + b\,\xi + c) + 2\,a\,\xi + b\right),$$

$$a = \frac{1}{2\gamma_1 \gamma_2};\ \ b = \frac{(\gamma_1+\gamma_2)}{(\gamma_1 \gamma_2)^2};\ \ c = \frac{(\gamma_1^2+\gamma_1 \gamma_2+\gamma_2^2)}{(\gamma_1 \gamma_2)^3};\ \ \gamma_{1,2} = m + \chi_{1,2}.$$

Here in the case of double sign $\mp$ or $\pm$, the first index corresponds to the upper sign while the second to the lower.

The photoresponse $V_{\text{in}}(\xi)$ is proportional to $T_e(\xi) - T_0$ (Eq. 10), the steepness of the superconducting transition at the operation point, and the bias current. This initial transient is modified by the readout electronics (cables, bias-T, amplifiers, and sampling oscilloscope) with the finite bandpass. Transient characteristic of the readout, which is the output voltage transient in response to the unit vertical voltage step at the input, can be sufficiently good described as

$$h(\xi) = \left(1 - e^{-2\sqrt{2}f_C \xi}\right) \cdot e^{-2\sqrt{2}f_S \xi}, \tag{12}$$

where $f_S$ and $f_C$ – are the lower and the upper frequencies of the bandpass. Knowing $V_{\text{in}}(\xi)$, one can compute voltage transient at the oscilloscope with the Duhamel integral as

$$V_{\text{out}}(\xi) = \int_0^\xi \dot{V}_{\text{in}}(\xi')\, h(\xi - \xi')\,\mathrm{d}\xi'. \tag{13}$$

We used the formalism described by Eqs. (9 -13) to fit voltage transients recorded by the oscilloscope. The ringing was simulated by adding a series of equidistant identically shaped pulses with decreasing magnitudes. The best fit curves are shown in Fig. 4 with solid lines. There are four independent parameters: the heat capacity ratio $c_e/c_{\text{ph}}$, the *e-ph* energy relaxation time, $\tau_{\text{EP}}$, the phonon escape time, $\tau_{\text{esc}}$, and the normalized pulse energy $P_0/c_e$. The latter changes only the magnitude of the transient and does not affect its shape. For each bridge, we fixed $\tau_{\text{EP}}$ at the value resulted from averaging over the electron states (Eq. 1) of the *e-ph* scattering rate, which was obtained by MC measurements (Table III). We also fixed phonon escape times at the values computed for each bridge in the framework of the acoustic mismatch model (for details see Sec. IV B). This leaves only one fitting parameter $c_e/c_{\text{ph}}$. The best fit values of $c_e/c_{\text{ph}}$, together with computed values of $\tau_{\text{esc}}$ are listed in Table III. Heat capacity ratios $c_e/c_{\text{ph}}$ scatter in the range of 0.1 – 0.83. These values agree reasonably well with the previously reported data. In section III C, we compare the best fit values of $c_e/c_{\text{ph}}$ with predictions of the Debye and Drude models.

### D. Photoresponse in the frequency domain

Frequency-domain measurements were done for K-series of NbN microbridges on sapphire substrates (Table II samples K1-K9). Film thickness varied from 3.2 to 33.2 nm. Data obtained with the frequency domain technique were partly reported in Refs. 15 and 40. The technique in detail was described in Ref. 15. Shortly, the microbridge was cooled down to an operating temperature within the resistive transition and biased by a small dc current. Beams of two continuous-wave near-infrared lasers (wavelength 850 nm) with the controllable difference between radiation frequencies were overlapped on the microbridge. The power of radiation that is absorbed by the microbridge alternates periodically at the beating frequency $f$ (the difference between frequencies of two lasers) and causes sinusoidal modulation of the electron temperature with the amplitude $\delta T_e(f)$. This leads to periodic sinusoidal variations in the photoresponse with the amplitude $\delta U(f) \sim \delta T_e(f)$. Oscillations in the photoresponse are amplified and controlled with a spectrum analyzer in the range of beating frequencies from 10 MHz to 10 GHz. Below we refer to the squared amplitude of these oscillations as the photoresponse magnitude $\delta P(f) \sim \delta U^2(f)$ which is expressed in decibels. A similar approach (AMAR) described in Ref. 32 differs only in radiation frequencies, which were in the sub-THz frequency range. The roll-off frequency $f_0$ in the dependence of the photoresponse magnitude on the beating frequency is the frequency at which the magnitude decreases to one half of its value at small frequencies $\frac{1}{2}\delta P(0)$. In the inset in Fig. 5 we show representative experimental data $\delta P(f)$ for the microbridge K-8 with the thickness 21.6 nm (open symbols) and the best fit (solid curve) obtained with the expression $\delta P(f) = \delta P(0)/(1 + f^2/f_0^2)$. For each microbridge operated at $T \approx T_C$, the roll-off frequencies were obtained from the best fit, and the response times were found as $\tau_\varepsilon = (2\pi f_0)^{-1}$. In Fig. 5 we plot the response time $\tau_\varepsilon$ as a function of the film thickness. The response time varies from 124 ps for the thinnest film to 421 ps for the thickest film. Generally, $\tau_\varepsilon$ decreases when $d$ decreases. However, the rate of the decrease is noticeably less for microbridges with smaller thicknesses.

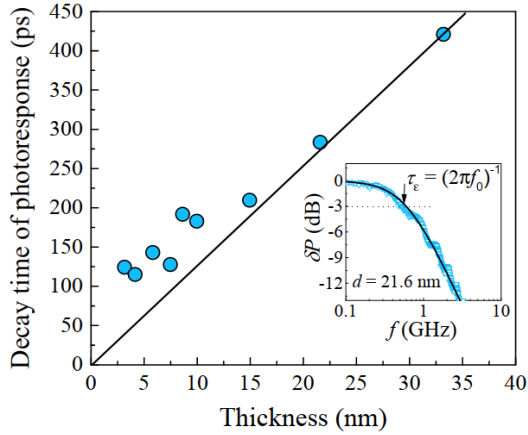

Fig. 5. Response time $\tau_\varepsilon$ vs thickness for NbN microbridges on sapphire substrates (symbols). The black line represents the linear fit $\tau_\varepsilon = 11.5\,d$ that corresponds to the computed phonon escape time versus film thickness. The inset shows a representative experimental $\delta P(f)$ curve (symbols) for the sample K-8. The black curve is the fit described in the text.

To obtain heat capacity ratios in the framework of the FDAM technique, we describe experimental response times, $\tau_\varepsilon$, with the 2-T model. We use Eq. (9) where we neglect dissipated Joule power and substitute periodic excitation in the form $P_{\mathrm{RF}}(t) = P_0 e^{-j2\pi f t}$ ($j = \sqrt{-1}$) for the pulse excitation. Here $P_0$ is the radiation power absorbed per unit area of the film. The solution for periodic excitation in the frequency domain was obtained by Perrin and Vanneste [16] and is given by

$$\delta T_e(f) = P_0 \frac{1}{d\,c_e} \frac{\tau_2 \tau_3}{\tau_1} \left| \frac{(1 + j2\pi f \tau_1)}{(1 + j2\pi f \tau_2)(1 + j2\pi f \tau_3)} \right|, \tag{14}$$

where characteristic times are $\tau_1 = (\Gamma_2 + \Gamma_3)^{-1}$ and $\tau_{2,3} = \chi_{1,2}^{-1}$ as defined in Eqs. (11) with $\Gamma_1 = \tau_{\mathrm{EP}}^{-1}$, $\Gamma_2 = \Gamma_1 \frac{c_e}{c_{\mathrm{ph}}}$, and $\Gamma_3 = \tau_{\mathrm{esc}}^{-1}$. The spectrum of the photoresponse (Eq. (14)) crucially depends on the heat capacity ratio $c_e/c_{ph}$. For instance, if $c_e \gg c_{ph}$ or $\tau_{EP} \gg \tau_{esc}$ inclusively, Eq. (14) reduces to $\delta T_e(f) \approx P_0 \tau_2 (d\, c_e)^{-1}(1 + (2\pi f)^2 \tau_2^2)^{-1/2}$, with $\tau_2 \approx \tau_{EP} + (c_e/c_{ph})\tau_{esc}$. Exactly this limiting case is valid for thin Nb films [70]. It was also used for thin NbN films in several works [20, 22]. However, for NbN the required inequalities are not satisfied. Indeed, the ratio $c_e/c_{ph}$ estimated from our measurements (Sec. III C) as well as the ratios obtained in Ref. 33 (for 5.5 nm NbN film at $T_C$) and in Ref. 21 all give $c_e < c_{ph}$. At the same time, even for the thinnest films $\tau_{EP} \lesssim \tau_{esc}$. This is why we used the full solution (Eq. 14) to fit our experimental data.

As it follows from MC measurements (Sec. III B, Table III) for films of the M-series, $\tau_{e-ph}$ as well as the exponent $n$ do not vary much with the degree of disorder. For films K1-K9 having a slightly less degree of disorder than films of the M-series we therefore expect close values of $n$ and close values of $\tau_{e-ph}$ at 10K. We assigned to all films of the K-series the mean values of the mass density ($\rho$ = 7.5 g/cm³), the phonon velocity ($u_t$ = 2.4×10³ m/s) and the electron mean free path ($l$ = 0.13 nm) found for films of the M-series except the most disordered film M-A853 (Table III, Table IV). Applying the SM formalism (Sec. IV A), we computed values of $\tau_{e-ph}$ at the actual transition temperatures for each film of the K-series. The values of $\tau_{e-ph}$ were further averaged according to Eq. (1) for the mean value $n$ = 3.85. These averaged values, $\tau_{EP}(T_C)$, fall into the range 0.33 – 0.13 ps. To get the theoretical dependence $\delta T_e(f)$, we used these fixed values of $\tau_{EP}(T_C)$ and set escape times at the values $\tau_{esc}$ [ps] = 11.5 $d$ [nm] computed in the framework of the acoustic mismatch model [5] (computational details for $\tau_{esc}$ are presented in Sec. IV B). We obtained the theoretical response times, $\tau_r$, within the 2-T model (Eq. 9) from Eq.(14) using the relation $\delta T_e^2(2\pi/\tau_r) = \frac{1}{2}\delta T_e^2(0)$. This approach leaves us the only one fitting parameter $c_e/c_{ph}$. The best fit values of $c_e/c_{ph}$, i.e. those providing the theoretical response times equal to the experimental response times ($\tau_r = \tau_\varepsilon$), vary monotonously from 0.1 for the thickest film with $d$ = 33.2 nm to 2.15 for the thinnest film with $d$ = 3.2 nm. We cross checked these results by applying the three-temperature (3-T) model [13]. Although the 3-T model differently accounts for phonon trapping as compared to the 2-T model, the best fit values of $c_e/c_{ph}$ were found to be very close to those provided by the 2-T model.

It is worth mentioning here that the accuracy of extracting best fit values is different for FDAM and TDP techniques. For the range of fitting parameters typical for NbN, the same variation in the fitting parameters results in similar changes in fitting curves. However, in the TDP transients this change is more pronounced in the area with the lowest noise around the peak, while in the FDAM spectra changes occur mostly beyond the roll-off frequency in the area with largest noise.

## IV. DATA ANALYSIS

Our results contain two important findings. First, the inelastic electron-phonon scattering rate depends on temperature as $1/\tau_{e-ph} \propto T^n$ with a weakly varying exponent $n \approx 3.2 \div 3.8$. In the next subsection, we analyze experimental $\tau_{e-ph}(T)$ dependences with the SM theory of electron-phonon interaction in dirty metal films [8]. Second, in the framework of the 2-T model, we obtained the best fit values for the ratio $c_e/c_{ph}$ for films with thicknesses in the range 3.2 – 33.2 nm. In section IV C, we compare these values with the predictions of the Drude and Debye models.

### A. Inelastic electron-phonon scattering time

According to the SM theory [8] the impact of disorder on the electron-phonon coupling is controlled by the product $q_T l$, where $q_T = k_B T/\hbar u$ is the wave vector of the thermal phonon and $u$ is the sound velocity. In a strongly disordered metal with $q_T l \ll 1$, the exponent, $n$, in the temperature dependence of the electron-phonon scattering rate $\tau_{e-ph}^{-1} \propto T^n$ is itself temperature dependent and can take any value between 2 and 4, depending on the degree of disorder and the property of elastic scatterers.

The inelastic scattering rate of an electron at the Fermi surface due to the interaction with longitudinal phonons (we use indices $l$ and $t$ to denote values associated with longitudinal and transverse phonon modes) is given by [8]

$$\tau_{e-ph(l)}^{-1} = \frac{7\pi\varsigma(3)}{2\hbar} \frac{\beta_l (k_B T)^3}{(p_F u_l)^2} F_l(q_{T(l)} l). \tag{15}$$

In the expression above $\varsigma(n)$ is the Riemann zeta function and $\beta$ is the dimensionless coupling constant. For both phonon modes it is given by $\beta_{l,t} = (2E_F/3)^2 (N(0)/(2\rho u_{l,t}^2))$. Here $E_F = p_F^2/(2m_e)$ is the Fermi energy, $p_F = N(0)\pi^2 \hbar^3/m_e$ is the Fermi momentum, $m_e$ is the electron mass, $\rho$ is the mass density and $u_{l,t}$ is the phonon velocity for a particular mode. The effect of disorder on the scattering rate is controlled by the integral $F_l(z) = \frac{2}{7\varsigma(3)} \int_0^{A_l} dx\, \Phi_l(xz)[N(x) + f(x)]x^2$, where $N(x)$ and $f(x)$ are Bose and Fermi distribution functions, and $\Phi_l(y) = \frac{2}{\pi}\left(\frac{y \arctan(x)}{y-\arctan(x)} - \frac{3}{y}k\right)$ is the Pippard function. The upper limit of the integral $F_l(z)$ is $A_{l,t} = (6\pi^2)^{1/3}(l/a)/z$, where $a$ is the size of the unit cell which is assumed for all films to be the same and equal to 0.44 nm. The parameter $1 \geq k \geq 0$ reveals the character of electron scatterers; $k = 1$ corresponds to scatterers vibrating together with the host lattice, $k = 0$ corresponds to the static (i.e. 'non-vibrating') scatterers such as heavy impurities and rigid boundaries. The inelastic electron scattering rate of an electron at the Fermi surface due to the interaction with transverse phonons is given by

$$\tau_{e-ph(t)}^{-1} = 3\pi^2 \frac{\beta_t (k_B T)^2}{(p_F u_t)(p_F l)} k F_t(q_{T(t)} l), \tag{16}$$

where $F_t(z) = \frac{4}{\pi^2} \int_0^{A_t} dx\, \Phi_t(xz)[N(x) + f(x)]x$, and the Pippard function $\Phi_t(x) = 1 + k \frac{3x - 3(x^2+1)\arctan(x)}{2x^3}$. The apparent electron-phonon scattering rate is the sum of the two rates $\tau_{e-ph}^{-1} = \tau_{e-ph(l)}^{-1} + \tau_{e-ph(t)}^{-1}$.

We fit our MC data (Sec. III B) using Eqs. (15-16). It turned out that the observed exponent in the temperature dependence of the scattering rate could only be reproduced with $k = 1$ for samples M-2559, M-A854, and M-A855. In the temperature range where our MC data were acquired, the scattering rate of electrons via transverse phonons dominates, and the parameter $u_l$ does not affect the result of simulations. We, therefore, excluded $u_l$ from the set of fit parameters and took it instantly twice as large as $u_t$. The relation $u_l = 2u_t$ is approximately valid for a large variety of materials. The remaining fit parameters are $l, \rho,$ and $u_t$. Their best fit values are listed in Table IV. We used the density of electron states computed from the data of transport measurements (Sec. III A) and the free electron mass (see Ref. [27] for verification) to obtain Fermi momentum and energy.

Table IV. Best fit values of the parameters in the SM theory.

| sample | $l$ (nm) | $\rho$ (g / cm$^3$) | $u_t$ (m/s) |
|---|---|---|---|
| M-2259 | 0.13 | 7.8 | $2.42 \times 10^3$ |
| M-A853 | - | 5.2 | $2.2 \times 10^3$ |
| M-A854 | 0.14 | 7.5 | $2.4 \times 10^3$ |
| M-A855 | 0.12 | 7.5 | $2.37 \times 10^3$ |

For all samples, the values of the electron mean free path $l$ are by a factor of two smaller than the values obtained by different groups [25, 27, 29] from Hall-effect measurements, and by a factor of 4 to 6 smaller than the values reported in Ref. 24 where they were computed as $(3D\tau)^{1/2}$ from the measured diffusion constant and the elastic-scattering time. The latter was obtained by means of spectral ellipsometry. The best fit values of the velocity of transverse phonons and the mass density deviate from those reported in Ref. 71 where for bulk hexagonal NbN these parameters were found 4.5×10³ m/s and 8.5 g/cm³, respectively. Diversely, $\rho$ and $u_t$ obtained as fitting parameters for NbN are similar to those for TaN [72]. This finding correlates with remarkable similarity in the superconducting properties of these two materials [73, 74]. A smaller mass density correlates with the excess nitrogen content with respect to the optimal stoichiometry while reduced sound velocity is most probably the result of granularity. For all films of the M-series, the product $q_T l \ll 1$. It can be presented as $q_T l = \alpha T$ where the coefficient $\alpha$ falls into the range 0.075±0.005 K⁻¹. Hence, the films of the M-series are strongly disordered with a very close degree of disorder.

Fitting the data for the film M-A853 with $k = 1$ gives an enormously large electron mean free path $l = 0.31$ nm that contradicts to other parameters ($D$ and $R_{SN}$). Although the exact reason is not clear, we have to note that using $k \approx 0.9$ results in a reasonably small $l$. In the SM theory, $k < 1$ corresponds to the presence of static scatterers. Since the grain boundaries are a kind of static, non-vibrating scatterers, experimental data can be qualitatively related to the enhanced granularity of the film M-A853 as compared to other films of the M-series. The Ioffe-Regel parameter estimated for this film is $k_F l \lesssim 1$ that may also indicate enhanced granularity [35-37].

### B. Phonon escape time

We use the acoustic mismatch model by Kaplan [5] to compute phonon transmission coefficients for metal/substrate interfaces NbN/SiO$_2$ and NbN/Al$_2$O$_3$. The model describes acoustic plane waves associated with different phonon modes, which propagate through the interface between two isotropic semi-infinite media with zero attenuation, and takes into account mode conversion and total reflection at the interface. For instance, an incident longitudinal phonon (wave) is reflected and transmitted as pairs of longitudinal and transverse phonons (waves). Reflection and transmission coefficients depend on the angle of incidence, $\theta$, propagation velocities of the modes in both media and the difference between their acoustic impedances $Z_{1,2,i} = u_{1,2,i} \rho_{1,2}$ which are the products of the mass density of the medium and the propagation velocity of the particular mode in this medium. Here indices 1 and 2 refer to the film and the substrate, the index i denotes the mode. Applying boundary conditions, which require continuity of the mechanical strain and stress at the interface, we found ratios of amplitudes of reflected and transmitted waves to the amplitude of the incident wave. Angle dependent transmission coefficients were defined separately for longitudinal, $\eta_{\theta l}$, and two transverse, $\eta_{\theta ts}$ and $\eta_{\theta tp}$, modes as the ratio of the total energy flux of all transmitted modes to the energy flux of the incident mode $P_i = \omega^2 Z_{1i} A_i^2 \cos\theta / 2$ where $A_i$ is the amplitude of the incident mode. Phonon escape time was defined for a flux of phonons propagating within a narrow solid angle at an incident angle $\theta$ and undergoing multiple sequential specular reflections at interfaces with the substrate and vacuum at the other side of the film. We define escape time for the mode as $\tau_{esc}(\theta)_i = P(t)(dP(t)/dt)^{-1}$ where $P(t)$ is the phonon flux remaining in the film. Right before the reflection with the number $q$ the relative amount of $P(t)$ is $(1 - \eta_{\theta i})^{q-1}$ and decreases after the reflection by the factor $1 - \eta_{\theta i}$. The time between two sequential reflections equals $2 d/(u_i \cos\theta)$ that results in the dimensionless rate of the decrease in the photon flux $\tau_{esc}(\theta)_i^{-1} = u_i \eta_{\theta i} \cos\theta /(2d)$. Integration over the solid angle gives the escape time per mode $\tau_{esc,i}^{-1} = u_i \eta_i /(4d)$ with the angle-averaged transmission coefficient for a particular mode $\eta_i = 2 \int_0^{\theta_{max,i}} \eta_{\theta i} \sin\theta \cos\theta \, d\theta$, where $\bar{\theta}_{max,i} = arcsin(u_{1,i}/u_{2,i})$ is the critical angle of total internal reflection for this mode. Since the decay rate of the photon energy through the particular mode is proportional to the heat capacity of the mode, which in turn is inversely proportional to the cube of the mode velocity, we found total weighted escape rate $\tau_{esc}^{-1} = \frac{\overline{\eta u}}{4d}$, where $\overline{\eta u} = \frac{\sum_i u_i^{-2} \eta_i}{\sum_i u_i^{-3}}$. Weighted values for the transmission coefficient and mode velocities were obtained in a similar way as $\bar{\eta} = \sum_i u_i^{-3}\eta_i / \sum_i u_i^{-3}$ and $\bar{u} = \sum_i u_i^{-2} / \sum_i u_i^{-3}$. We have to note here that although $\overline{\eta u} \neq \bar{\eta}\,\bar{u}$, the difference between two sides of this inequality for studied interfaces remains less than ten per cent. Values of mass densities and sound velocities for substrates we took from Ref. 5. For NbN we used values obtained via fitting procedure in the framework of the SM theory (Sec. IV A) and retained the assumption that the velocity of longitudinal phonons is twice as large as that of transverse phonons. We found for NbN/SiO$_2$ interface $\bar{\eta}$ = 0.28, $\bar{u}_1$= 2.54 × 10³ m/s, $\bar{u}_2$= 4.35 × 10³ m/s and for NbN/Al$_2$O$_3$ interface $\bar{\eta}$ = 0.12, $\bar{u}_2$= 6.87 × 10³ m/s. For NbN/SiO$_2$ we obtained $\tau_{esc}$[ps] =5.2 $d$ [nm] and for NbN/Al$_2$O$_3$ $\tau_{esc}$[ps] =11.5 $d$ [nm], these values of phonon escape times are used in Sec. III C, respectively.

Computations including all three modes showed that for both studied interfaces the energy is dominantly transferred via transverse modes $\eta_\theta \approx \eta_{\theta ts} + \eta_{\theta tp}$ and that $\eta_\theta$ decreases very slowly with the angle until the angle of total internal reflection $\bar{\theta}_{max} = arcsin(\bar{u}_1/\bar{u}_2)$.

### C. Discussion

When compared at the same temperatures, the best fit values of the heat capacity ratios for thinnest films of K-series is approximately 25% larger than the ratios obtained for films of M-series. This observation agrees with the Drude and the Debye models for electrons and phonons, respectively, if one takes into account temperature dependences of heat capacities, $c_e \propto N(0)\,T$ and $c_{ph} \propto T^3$, and the difference between densities of electron states (Table II) for films of the K and M series.

Let us now compare the absolute values of heat capacity ratios predicted by the Drude and Debye models with the ratios obtained experimentally as best fits in the framework of the 2-T model. The Drude model predicts for electrons the heat capacity $c_e = \frac{\pi^2 k_B^2}{3} N(0)\,T$. For phonons, the 3-d Debye model predicts the heat capacity $c_{ph} = \frac{2\pi^2 k_B}{15}\left(\frac{k_B T}{\hbar}\right)^3 \left[2\left(\frac{1}{u_{1t}}\right)^3 + \left(\frac{1}{u_{1l}}\right)^3\right]$. Taking $N(0)$ from Table II, and $u_{1t} = u_{1l}/2 = 2.4\times10^3$ m sec$^{-1}$ (Table IV) we obtained $c_e$ and $c_{ph}$ at the actual critical temperatures of each studied sample. Computed model ratios $c_e/c_{ph}$ for very thin films are less than the values obtained via best fits of FDAM and TDP data. Let us assume that the electron heat capacity is described quantitatively well by the Drude model. Then the phonon heat capacity in thin films is less than the Debye model predicts. The values of $c_{ph}$ computed with the Debye model are shown in Fig. 6, together with the best fit values. To obtain the best fit values for $c_{ph}$, we assigned to films of the K-series averaged values of $N(0)$ and $n$. However, as it is seen in Table III and Ref.[24], these parameters vary with the thickness. Error bars in the right graph of Fig. 6 show expected uncertainties in the phonon heat capacities. With the decrease of the film thickness, the difference between the values of $c_{ph}$ obtained via best fits in the framework of the 2-T model and the values predicted by the Debye model increases.

We further estimate the Debye temperature of our films in the framework of the 3-dimensional Debye model as $T_D = \hbar(6\pi^2)^{1/3}\bar{u}_1/(k_B a)$. Assuming $a$ = 0.44 and $\bar{u}_1$ = 2.54 × 10³ m/s (Subsection IV B), we found $T_D$ = 172 K. Such value is typical for Debye temperatures reported for similar films [27] and is a few times less than the values reported for bulk NbN. The reduction of the Debye temperature is usually denoted as "phonon softening" caused by granularity and weakening of ion bonds on film surfaces [75].

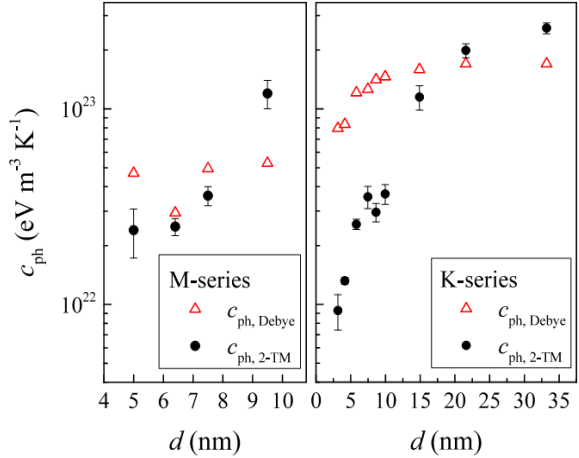

Fig. 6. Phonon heat capacities vs film thickness for films of the M-series (left graph) and the K-series (right graph) in the semi-logarithmic scale. Values $c_{ph,\,Debye}$ (open symbols) were computed with the 3-d Debye model and phonon velocities found in section IV A. Values $c_{ph,\,2\text{-TM}}$ were extracted from the best fit ratios $c_e/c_{ph}$ with values of $c_e$ predicted by the Drude model. Error bars in the right graph show the impact of variations in $N(0)$ and the exponent $n$ between films of the K-series. Error bars in the left graph correspond to uncertainties in the best fit values of $c_e/c_{ph}$.

The differences between phonon heat capacities obtained as the best fits with the 2-T model and computed in the framework of the Debye model is most pronounced for thin films. We attribute this difference to changes in the phonon spectrum. There are at least two effects that may cause a decrease in the phonon heat capacity in thin films. One is the depletion of the transverse phonon modes in thin films specifically for the mode with polarization along the normal to the film [75]. Another one is an increase in the phonon wavelength at low temperatures. As a consequence, in thin films at low temperatures, the phonon wavelength becomes comparable or even larger than the film thickness. This effect eliminates low-energy phonons propagating at small angles $\theta$ and hence destroys isotropy of the phonon spectrum and reduces phonon density of states. Using the 3-d density of states from the Debye model and the average kinematic velocity for phonons $u = 4u_{1t}/3$ we arrive at $\lambda \approx 2\pi\hbar u/(k_B T) \approx 12$ nm at 13 K. This value is four times the thickness of the thinnest film in the K-series. The condition $\lambda \geq d$ restricts available directions of the phonon wave vectors most efficiently within the cone with plane angle $\bar{\theta}_{max}$ around the normal to the interface where $\eta_\theta > 0$ and where phonons can only escape from the film. The reduction in the phonon spectrum emitted perpendicularly to the film/substrate interface with the decrease of the film thickness was observed and modeled in [76]. The authors showed that in the restricted direction, the phonon spectrum is modified. Phonon states with small frequencies are forbidden that resulted in discrete, sharp steps in the number of excited phonons.

## V. CONCLUSION

We have studied inelastic scattering and energy relaxation of electrons by means of magnetoconductance and photoresponse, respectively, in a series of superconducting NbN films on Si/SiO$_2$ and Al$_2$O$_3$ substrates with thicknesses in the range from 3 to 33 nm. Our main results are:

(a) In studied NbN films in the temperature range from 10 to 30 K, the inelastic electron-phonon scattering rate defined by magnetoconductance technique depends on temperature as $1/\tau_{e-ph} \propto T^n$ with the exponent $n \approx 3.2 \div 3.8$. The magnitude of $\tau_{e-ph}$ at 10 K falls into the range 11.9 - 17.5 ps. In this temperature range, the films are strongly disordered. They are characterized by close values of the product $q_T l =$ $\alpha T \ll 1$, which controls the impact of disorder on the electron-phonon coupling. The coefficient $\alpha$ falls into the range 0.075±0.005 K$^{-1}$, while the Ioffe-Regel parameter, $k_\text{F}l$, varies in the range from 0.88 to 1.22.

(b) The Debye temperature in our films is noticeably smaller than the Debye temperature of bulk NbN material. We attribute this to phonon softening caused by granularity and weakening of ion bonds at film surfaces.

(c) Experimental photoresponse data for thicker films are described reasonably well in the framework of the 3-d Debye model and the 2-T model with the film parameters extracted from magnetoconductance measurements. Photoresponse of thinner films can only be described with a heat capacity of phonons smaller than the Debye model predicts. We attribute this finding to the reduced density of phonon states in thin films with thicknesses comparable or smaller than the wavelength of thermal phonons.

**ACKNOWLEDGMENTS**

M.S. acknowledges support by the Helmholtz Research School on Security Technologies, M.M. acknowledges support by the Russian Foundation for Basic Research (project No. 19-32-90083). The work was partly supported by the German Federal Ministry of Education and Research (Programm ERA.Net RUS Plus, Project ID: 88) and the Russian Science Foundation (project No. 17-72-30036). The authors would like to thank D. Henrich for providing raw experimental data.